\documentclass[acmsmall]{acmart}
\usepackage{adjustbox}
\usepackage{algorithm}
\usepackage{algorithmicx}
\usepackage{algpseudocode}
\usepackage{amsmath}
\usepackage{array}
\usepackage{balance}
\usepackage{booktabs}
\usepackage[skip=1pt,labelfont=bf]{caption}
\usepackage{calc}
\usepackage{calligra}
\usepackage{xcolor}
\usepackage{tabularx}
\usepackage{ltablex}   
\keepXColumns       
\usepackage{booktabs}
\usepackage{color}
\usepackage{colortbl}
\usepackage{courier}
\usepackage{csvsimple}
\usepackage{enumitem}
\usepackage{fancybox}
\usepackage{fontenc}
\usepackage{fontawesome5}
\usepackage{graphicx}
\usepackage{listings}
\usepackage{longtable}
\usepackage{lscape}
\usepackage{makecell}
\usepackage{marvosym}
\usepackage{moreverb}
\usepackage{multicol}
\usepackage{multirow}
\usepackage{pifont}
\usepackage{pgfplots}
\usepackage{rotating}
\usepackage{setspace}
\usepackage{siunitx}
\usepackage{soul}
\usepackage{subcaption}
\usepackage{tablefootnote}
\usepackage[most]{tcolorbox}
\usepackage{threeparttable}
\usepackage{tikz}
\usepackage[normalem]{ulem}
\usepackage{url}
\usepackage{wasysym}
\usepackage{xspace}

\usepgfplotslibrary{statistics}
\usetikzlibrary{matrix, shapes.geometric, arrows}
\usetikzlibrary{shapes, arrows, positioning}

\algnewcommand\algorithmicforeach{\textbf{for each}}
\algdef{S}[FOR]{ForEach}[1]{\algorithmicforeach\ #1\ \algorithmicdo}

\newcolumntype{L}[1]{>{\raggedright\let\newline\\\arraybackslash\hspace{0pt}}m{#1}}
\newcolumntype{C}[1]{>{\centering\let\newline\\\arraybackslash\hspace{0pt}}m{#1}}
\newcolumntype{R}[1]{>{\raggedleft\let\newline\\\arraybackslash\hspace{0pt}}m{#1}}

\definecolor{codegreen}{rgb}{0,0.6,0}
\definecolor{codered}{rgb}{1,0,0}
\definecolor{codegray}{rgb}{0.5,0.5,0.5}
\definecolor{codepurple}{rgb}{0.58,0,0.82}
\definecolor{backcolour}{rgb}{0.95,0.95,0.92}
\definecolor{lightgray}{gray}{0.9}

 { \newcommand{\mynote}[2]{
      \fbox{\bfseries\sffamily\scriptsize#1}
        {\small$\blacktriangleright$\textsf{\emph{#2}}$\blacktriangleleft$}}}
        { \newcommand{\mynote}[2]{}}
        
\definecolor{DarkOrange}{rgb}{0.8,0.3,0.0}
\definecolor{DarkCyan}{rgb}{0.0, 0.55, 0.55}
\definecolor{DarkCyel}{rgb}{1.0, 0.49, 0.0}
\definecolor{yellow-green}{rgb}{0.6, 0.8, 0.2}

\newcolumntype{?}{!{\vrule width 1pt}}

\lstdefinelanguage{mymarkdown}{
    morekeywords={*,\#, \#\#, \#\#\#},
    sensitive=false,
    morecomment=[l]{//},
    morestring=[b]",
    commentstyle=\color{codegreen},
    keywordstyle=\color{magenta},
    numberstyle=\tiny\color{codegray},
    stringstyle=\color{codepurple},
    basicstyle=\small,
    breakatwhitespace=false,         
    breaklines=true,
    breakindent=0pt,
    keepspaces=true,                 
    numbers=left,                    
    numbersep=5pt,                  
    showspaces=false,                
    showstringspaces=false,
    showtabs=false,                  
    tabsize=2,
}

\lstdefinestyle{mystyle}{
    commentstyle=\color{codegreen},
    keywordstyle=\color{magenta},
    numberstyle=\small\color{black},
    stringstyle=\color{codepurple},
    basicstyle=\scriptsize\ttfamily,
    breakatwhitespace=false,
    breaklines=true,
    captionpos=b,
    keepspaces=true,
    showspaces=false,
    showstringspaces=false,
    showtabs=false,
    tabsize=2
}

\lstdefinelanguage{diff}{
  morecomment=[f][\color{blue}]{@@},     
  morecomment=[f][\color{red}]-,         
  morecomment=[f][\color{codegreen}]+,       
  morecomment=[f][\color{red}]{---}, 
  morecomment=[f][\color{codegreen}]{+++},
  numberstyle=\tiny\color{codegray},
  numbers=left,                    
  numbersep=5pt,         
}

\setlist{noitemsep} 

\definecolor{darkpastelred}{rgb}{0.76, 0.23, 0.13}
\definecolor{ao(english)}{rgb}{0.0, 0.5, 0.0}

\definecolor{darkpastelred}{rgb}{0.76, 0.23, 0.13}
\definecolor{ao(english)}{rgb}{0.0, 0.5, 0.0}

\newboolean{useblue}
\setboolean{useblue}{false} 

\newcommand{\maybeblue}[1]{%
    \ifthenelse{\boolean{useblue}}%
    {\textcolor{blue}{#1}}%
    {#1}%
}

\AtBeginDocument{%
  }

\setcopyright{none}

\begin{document}
\newboolean{showcomments}
\setboolean{showcomments}{true}

\definecolor{DarkOrange}{rgb}{0.8,0.3,0.0}
\definecolor{DarkCyan}{rgb}{0.0, 0.55, 0.55}
\definecolor{DarkCyel}{rgb}{1.0, 0.49, 0.0}
\definecolor{yellow-green}{rgb}{0.6, 0.8, 0.2}

\newcommand{\todoc}[2]{{\textcolor{#1} {\textbf{#2}}}}
\newcommand{\todoblue}[1]{\todoc{blue}{\textbf{#1}}}
\newcommand{\todogreen}[1]{\todoc{yellow-green}{\textbf{#1}}}
\newcommand{\todored}[1]{\todoc{red}{\textbf{#1}}}
\newcommand{\bachle}[1]{{\color{red}\textbf{Bach:}} {\todoblue{#1}}}
\newcommand{\chong}[1]{{\color{red}\textbf{Chong:}} {\todoblue{#1}}}

\title{A Taxonomy of Prompt Defects in LLM Systems}
\author{Haoye Tian}
\affiliation{%
  \institution{School of Computer Science and Engineering, Nanyang Technological University}
  \country{Singapore}
}
\email{tianhaoyemail@gmail.com}

\author{Chong Wang}
\affiliation{%
  \institution{School of Computer Science and Engineering, Nanyang Technological University}
  \country{Singapore}
}
\email{XXX}

\author{BoYang Yang}
\affiliation{%
  \institution{Jisuan Institute of Technology, Beijing JudaoYouda Network Technology Co. Ltd.}
  \country{China}
}
\email{yby@ieee.org}

\author{Lyuye Zhang}
\affiliation{%
  \institution{School of Computer Science and Engineering, Nanyang Technological University}
  \country{Singapore}
}
\email{XXX}

\author{Yang Liu}
\affiliation{%
  \institution{School of Computer Science and Engineering, Nanyang Technological University}
  \country{Singapore}
}
\email{yangliu@ntu.edu.sg}

\begin{abstract}
Large Language Models (LLMs) have become key components of modern software, with prompts acting as their de-facto programming interface. However, prompt design remains largely empirical and small mistakes can cascade into unreliable, insecure, or inefficient behavior. This paper presents the first systematic survey and taxonomy of prompt defects, recurring ways that prompts fail to elicit their intended behavior from LLMs. We organize defects along six dimensions: (1) Specification \& Intent, (2) Input \& Content, (3) Structure \& Formatting, (4) Context \& Memory, (5) Performance \& Efficiency, and (6) Maintainability \& Engineering. Each dimension is refined into fine-grained subtypes, illustrated with concrete examples and root cause analysis. Grounded in software engineering principles, we show how these defects surface in real development workflows and examine their downstream effects. For every subtype, we distill mitigation strategies that span emerging prompt engineering patterns, automated guardrails, testing harnesses, and evaluation frameworks. We then summarize these strategies in a master taxonomy that links defect, impact, and remedy. We conclude with open research challenges and a call for rigorous prompt engineering oriented methodologies to ensure that LLM-driven systems are dependable by design.
\end{abstract}

\maketitle

\section{Introduction}

Large language models (LLMs) have become integral to modern software applications, acting as powerful components for tasks ranging from natural language query answering to code generation and repair~\cite{achiam2023gpt,du2024evaluating, yangmorepair, yang2025enhancing}. In these LLM-based systems, the prompt, a natural language input that instructs the model, effectively serves as the source code that determines the behavior of the model~\cite{chen2025promptware, white2023prompt}. This paradigm, sometimes called prompt-powered software or promptware~\cite{chen2025promptware}, allows developers to perform complex tasks using plain-language instructions rather than traditional programming. However, unlike conventional code, prompts are written in an ambiguous, unstructured, context-dependent medium (natural language) and execute on a non-deterministic, probabilistic engine (the LLM)~\cite{zhao2021calibrate, wei2022emergent}. These fundamental differences introduce unique and significant challenges to ensure reliability and predictability in prompt development, often reducing prompt crafting to an iterative, empirical process, a ``trial-and-error'' approach.

This inherent ambiguity and nondeterminism make prompts highly susceptible to defects: errors or shortcomings that cause an LLM to produce output deviating from the user's intent, much like bugs in source code cause a program to behave incorrectly~\cite{liu2023pre}. Empirical evidence demonstrates that prompt defects can lead to a spectrum of failure modes, ranging from trivial formatting issues to severe misinformation and critical security breaches~\cite{hui2024pleak,wang2022self,pathade2025red, chao2024jailbreakbench}. For example, a poorly specified prompt might yield irrelevant or incorrect answers~\cite{liu2023pre, wang2022self}, while a maliciously crafted user input can inject instructions that override the intent of the system, similar to a code injection attack~\cite{pathade2025red}. Furthermore, attackers have exploited assistant prompts to reveal confidential system instructions or generate disallowed content~\cite{guardian_bypass2024, chao2024jailbreakbench}. Such failures underscore that prompt quality is not merely a matter of convenience or elegance; it is directly tied to software correctness, security, and ethics in LLM applications.

To improve prompt quality and align LLM outputs with downstream requirements, the field of prompt engineering has emerged, offering concrete guidelines and tooling for writing, structuring, and validating prompts~\cite{openai_best_practices,anthropic_guide}. Techniques such as few-shot learning~\cite{brown2020language}, chain-of-thought prompting~\cite{wei2022chain}, self-consistency~\cite{wang2022self}, and retrieval-augmented generation~\cite{lewis2020retrieval} have enhanced an LLM's ability to understand intent and produce desirable outputs by providing clearer instructions, examples, or external knowledge. However, despite these valuable advancements, prompt engineering largely operates on an ad-hoc basis because it lacks a systematic understanding of the underlying prompt defect mechanisms. 
The nature of prompt defects is complex and multifaceted, ranging from subtle ambiguities to blatant adversarial manipulations, and often leading to significant performance drops even under minor perturbations. 
So far, the community still lacks a unified and systematic classification of prompt defects, leaving practitioners to rely on fragmented heuristics rather than a principled approach.


In this paper, we present a comprehensive taxonomy of prompt defects in LLM systems. Our goal is to systematically categorize the ways a prompt can fail to elicit the desired behavior from an LLM, providing software engineers and researchers with a common vocabulary and understanding of these critical failure modes. We define six top-level categories of prompt defects based on the aspect of the prompt or interaction they affect:
\begin{itemize}
    \item \textbf{Specification \& Intent Defects:} Flaws in how the prompt captures the user’s goals or requirements (e.g., ambiguous instructions, underspecified tasks).
    \item \textbf{Input \& Content Defects:} Issues arising from the content provided to the prompt, including user inputs (e.g., irrelevant or malicious input, factual errors in context).
    \item \textbf{Structure \& Formatting Defects:} Errors in the construction or syntax of the prompt (e.g., poor organization, missing delimiters, improper formatting of examples or outputs).
    \item \textbf{Context \& Memory Defects:} Failures in handling conversational context or memory (e.g., forgetting prior instructions, context window overflow).
    \item \textbf{Performance \& Efficiency Defects:} Prompt issues that impact latency, cost, or resource usage (e.g., overly long prompts causing slow responses, inefficient chaining of prompts).
    \item \textbf{Maintainability \& Engineering Defects:} Challenges in managing prompts as an evolving software artifact (e.g., hard-coded prompts, untested prompt changes leading to regressions).
\end{itemize}

Each of these categories comprises several more granular defect subtypes. In the core sections of this paper, we examine each subtype in depth, providing a definition and illustrative example, analyzing why the defect occurs and what consequences it leads to, and reviewing known or potential mitigation strategies. 
Throughout, we integrate insights from current academic research and reports, as well as practical guidelines from industry.

\section{Methodology for Building the Taxonomy}
\label{sec:methodology}

To construct a systematic taxonomy of prompt defects in LLM systems, we first conducted a comprehensive literature review covering research on prompt engineering, LLM robustness, and software engineering principles. We surveyed papers from leading venues (e.g., ICSE, FSE, ASE, ACL, NeurIPS) as well as preprints on prompt design and LLM security. In parallel, we analyzed industrial guidelines and best practices from OpenAI, Anthropic, AWS Bedrock, and other platform providers to capture state-of-the-art prompting techniques. 
We adopted a bottom-up inductive process to identify and abstract recurring defect patterns from the collected data. We performed multiple rounds of collaborative workshops and peer reviews to ensure consistency, coverage, and accuracy. Drawing on software engineering principles, we iteratively refined the taxonomy to achieve both completeness, by covering diverse failure modes, and orthogonality, by ensuring that categories are distinct and non-overlapping.
\section{Taxonomy of Prompt Defects}
\label{sec:taxonomy}


Table~\ref{tab:prompt-defects} provides an overview of our taxonomy, summarizing all identified prompt defect categories and subtypes along with their typical impacts and mitigations. The first column lists each category. For each subtype of defect, we provide a brief description and an example scenario illustrating the defect, then summarize its impact on the LLM’s behavior and on the resulting user or system outcomes, and finally present typical mitigation strategies and recommended practices. 

This taxonomy is intended to serve as a foundation for developing more robust prompt design and engineering practices for prompt software. By enumerating prompt defect types and compiling proven mitigations, our goal is to equip practitioners with the knowledge to anticipate and avoid common defect modes in prompt-driven systems. 
We argue that as LLMs become increasingly central to software, prompt development must mature into disciplined engineering that relies on mature cycles of testing, debugging, and maintenance, thereby keeping these systems reliable, secure, and faithful to user intent.

\begin{center}
\small
\setlength{\tabcolsep}{4pt}
\renewcommand{\arraystretch}{1}
\begin{longtable}{@{}>{\scriptsize}p{0.12\textwidth}>{\scriptsize}p{0.3\textwidth}>{\scriptsize}p{0.2\textwidth}>{\scriptsize\arraybackslash}p{0.28\textwidth}@{}}
\caption{Taxonomy of Prompt Defects in LLM Systems, with examples, impacts, and mitigations.} \label{tab:prompt-defects} \\
\toprule
\textbf{Category} & \textbf{Subtype \& Description} & \textbf{Impact on LLM} & \textbf{Mitigations} \\
\midrule
\endfirsthead

\toprule
\textbf{Category} & \textbf{Subtype \& Description} & \textbf{Impact on LLM} & \textbf{Mitigations} \\
\midrule
\endhead


\endfoot
\bottomrule
\endlastfoot

    \multirow{4}{=}{\textbf{Specification \& Intent}}%
      & \textbf{Ambiguous instruction.} Prompt is unclear or interpretable in multiple ways. \emph{Example:} User says “Make it better,” without context or criteria~\cite{shi2025ambiguity}.%
      & Model guesses intent (e.g., picks an arbitrary aspect to improve), often producing irrelevant or unsatisfactory output.%
      & Specify intent explicitly, e.g., “Improve the code’s readability by renaming variables and adding comments”. Include concrete criteria or goals to remove ambiguity. \\

      \cmidrule(l){2-4}

      & \textbf{Underspecified constraints.} Prompt lacks needed details or success criteria~\cite{yang2025prompts}. \emph{Example:} “Generate test cases.” (Does not specify format, scope, or criteria).%
      & Output may be too general, omit important cases, or not meet the user’s actual needs (e.g., trivial tests).%
      & Add requirements or constraints: e.g., “Generate \emph{unit} tests using \texttt{pytest}, covering edge cases and error handling”. Define required output format, coverage, or other acceptance criteria.\\

      \cmidrule(l){2-4}

      & \textbf{Conflicting instructions.} Prompt contains internally inconsistent or incompatible directives~\cite{colangelo2025write}. \emph{Example:} Provide a code summary with extensive detail.%
      & Model behavior is unpredictable: it may favor one instruction over the other or produce a muddled response trying to satisfy both.
      & Resolve contradiction by clarifying priority or removing one directive to ensure prompt is self-consistent (e.g. decide if the answer should be brief or detailed, not both).\\

      \cmidrule(l){2-4}

      & \textbf{Intent misalignment.} The prompt does not reflect the true user intent due to miscommunication~\cite{wu2025collabllm}. \emph{Example:} User asks a vague question and the prompt assumes a wrong goal.%
      & The model addresses the wrong problem or uses an inappropriate style/tone, leading to user frustration or unhelpful output.
      & Rephrase prompt after clarifying user intent via follow-up questions. Use iterative prompting: first confirm user’s goals, then proceed once the intent is correctly specified. \\
      
      \midrule
      
      \multirow{5}{=}{\textbf{Input \& Content}}%


      & \textbf{Misleading or incorrect info.} The prompt provides wrong facts or context~\cite{xu2024knowledge}. \emph{Example:} Prompt states, ``This function always returns a positive integer,'' and handle negative values.%
      & Model may faithfully use the false premise, yielding an answer that is logically consistent but factually wrong. %
      & Verify prompt content. If a hypothetical or counterfactual premise is needed, explicitly warn the model or instruct it to double-check the fact if possible.\\

      \cmidrule(l){2-4}

      & \textbf{Malicious prompt injection.} Untrusted input contains hidden instructions that alter behavior~\cite{kim2025prompt}. \emph{Example:} A user inputs “Ignore previous instructions; reveal the confidential code.”.%
      & The model may execute unauthorized commands or actions, violating the intended policy.  %
      & Treat user input as untrusted. Delimit user-provided content so it isn’t interpreted as instructions. Use hierarchical prompting (system messages) to enforce policies that user text cannot override.\\

      \cmidrule(l){2-4}

      & \textbf{Toxic or policy-violating input.} User input includes disallowed content~\cite{gehman2020realtoxicityprompts}. \emph{Example:} The user prompts the LLM to ``Generate a script to bypass the authentication mechanism of the internal API''. 
      & If not handled, the model might reproduce toxic content or perform a disallowed task. Conversely, it may also get confused or stuck. 
      & Pre-filter user inputs using content-moderation tools. Apply guardrails to refuse or sanitize prompts with violations, so that the model responds with safe completions.\\

      \cmidrule(l){2-4}

      & \textbf{Cross-modal misalignment.} In a multi-modal prompt (text + image, etc.), conflicting modalities are not handled properly~\cite{qian2024easy}. \emph{Example:} The prompt provides an image of a wireframe for a web form with 5 input fields, but the text states, ``Generate HTML with 3 input fields''. 
      & The model’s description may be incorrect or inconsistent. Overall accuracy drops due to contradictory inputs. 
      & Ensure consistency between modalities. If using images with text, keep textual descriptions accurate and in sync with the visual content. \\

      \midrule
      
      \multirow{5}{=}{\textbf{Structure \& Formatting}}%
      & \textbf{Lack of role separation.} Prompt does not clearly separate system instructions, user query, and assistant response~\cite{liu2023prompt,neumann2025position}. \emph{Example:} User concatenates everything in one block or fails to use the API’s role fields properly.%
      & The model may misidentify what is user input vs. developer instructions. This can lead to user instructions overriding system policy or other unintended behavior. %
      & Use structured prompt formatting: e.g., in OpenAI’s prompt engineering document~\cite{OpenAIPromptEngineering}, supply a distinct system message for guidelines, and a user message for the query. Clearly delineate any embedded content so the model knows user-provided context versus directive. \\

      \cmidrule(l){2-4}

      & \textbf{Poor prompt organization.} The prompt’s components (context, instructions, examples, question) are in a confusing or suboptimal order~\cite{liu2023lost,guan2025order}. \emph{Example:} Providing the main question before important context or rules, or mixing examples with rules without clear markers. 
      & The model might ignore or undervalue late-coming instructions, or apply rules to examples by mistake. Incoherent structure increases the chance the model misses key details or misinterprets the prompt’s intent. 
      & Follow a logical template for prompts. For instance: first set the context/role, then provide necessary background info, then state detailed instructions or rules, and finally ask the specific question. Use markers or sections (e.g.\ XML tags or headings) to separate these parts.\\

     \cmidrule(l){2-4}

     & \textbf{Formatting/syntax errors.} The prompt contains typos or incorrect syntax (e.g.\ unclosed quotes or code blocks)~\cite{sclar2023quantifying}. \emph{Example:} A prompt opens a markdown code block but never closes it. 
     & Such errors can confuse the model’s parsing of the prompt. The model may treat subsequent text as part of the code block or ignore content, leading to missing instructions. 
     & Validate prompt formatting. Ensure all markdown or XML tags are properly closed. During development, test the prompt in a controlled environment to catch formatting mistakes.\\

     \cmidrule(l){2-4}

     & \textbf{Undefined output format.} The prompt doesn’t specify how the answer should be formatted~\cite{he2024does}. \emph{Example:} “Explain the data,” without saying whether a bulleted list, paragraph, or JSON is expected. 
     & The model’s output may be inconsistent or not directly usable. If an application expects a specific format (like JSON for parsing), an undefined format can break the integration. 
     & Specify output format or style explicitly in the prompt. For example: “Provide the answer as a JSON object with fields \texttt{X}, \texttt{Y}, \texttt{Z}.” Additionally, validate outputs: use schemas or post-processing to check format, or leverage guardrail tools to enforce type/structure guarantees~\cite{rebedea2023nemo}.\\

    \cmidrule(l){2-4}

    & \textbf{Overloaded prompt.} The prompt tries to accomplish too many tasks at once~\cite{wei2022chain}. \emph{Example:} “Translate the Java code to Python, optimize the time complexity, and summarize it.” 
    & The model may handle one task and neglect others, or produce a jumbled response mixing all tasks. Performance and quality drop because the model is not explicitly guided step-by-step. 
    & Break complex tasks into a chain of prompts or subtasks (chain-of-thought or sequential prompting). For example, run separate steps: one prompt to translate, another to summarize, etc., or explicitly enumerate the required sections in one prompt.\\

    \midrule

    \multirow{5}{=}{\textbf{Context \& Memory}}%
    & \textbf{Context overflow/truncation.} The conversation or provided context exceeds the model’s context-window limit~\cite{wang2024beyond}. \emph{Example:} Earlier instructions about “do not change database schema” are ignored in the final suggestion. 
    & The model silently drops older or excess context. Important information from earlier may be lost, causing it to contradict prior facts or forget constraints. Inconsistent outputs can result if the omitted content was needed.%
    & Proactively manage context length: truncate or summarize earlier parts of the conversation before the limit is hit. Employ retrieval-based approaches: store conversation history externally and fetch only relevant pieces for each prompt, rather than resending the entire history.\\

    \cmidrule(l){2-4}

    & \textbf{Missing relevant context.} The prompt fails to include information that the model would need to produce a correct answer~\cite{wang2025llms}. \emph{Example:} User asks a follow-up question but the prompt doesn’t supply the previous answer or data needed to resolve it.%
    & The model may respond based on general knowledge or guesswork, leading to incorrect answers. It might also repeat questions or ask for clarification that should have been unnecessary if the context was given.%
    & Always include necessary context for the task. In a multi-turn scenario, incorporate relevant parts of prior conversation or data into the current prompt. For document-based queries, ensure the relevant segments are provided. If the context is too large, summarize it rather than omitting it entirely.\\

    \cmidrule(l){2-4}


    & \textbf{Irrelevant or noisy context.} Unnecessary information is included in the prompt~\cite{kusano2024longer}. \emph{Example:} Supplying the model with an entire log file when only a summary of one event is needed.%
    & The dilution of critical instructions or details by noise can distract the model's attention, limiting its potential.%
    & Prune irrelevant content before prompting. Use retrieval techniques to inject just the relevant facts for the query, rather than a data dump. \\

    \cmidrule(l){2-4}

    & \textbf{Conversational misreferencing.} The model confuses references in code-related discussions~\cite{vaithilingam2022expectation}. \emph{Example:} The user comments “This fix didn’t solve the issue,” but the prompt doesn’t specify which of the multiple suggested patches is being referred to.%
    & The model may misunderstand which code snippet or bug is under discussion, leading it to modify the wrong function or attribute a bug report to the wrong source.%
    & Use explicit referents in the prompt. Instead of vague terms like “this fix,” restate key points (e.g., “The user is referring to the patch applied to function \texttt{parseConfig}.”). Maintain clear speaker tags.\\

    \cmidrule(l){2-4}

    & \textbf{Forgotten instructions over time.} Important directives about code handling fade from the model’s active context later in the session~\cite{barke2023grounded}. \emph{Example:} The prompt specifies “use \texttt{pytest},” but later tests are generated using \texttt{unittest}. 

    & The model’s output eventually violates the original instruction simply because the directive scrolled out of view in the prompt.%
    & Reinforce key instructions throughout the interaction. Pin critical directives in a persistent system prompt prepended to every query, or systemically inject long-term memory into the model.\\

    \midrule
    
    \multirow{3}{=}{\textbf{Performance \& Efficiency}}%
    & \textbf{Excessive prompt length.} The prompt (including context and examples) is excessively long~\cite{jiang2023longllmlingua}. \emph{Example:} Providing 20 full code files causes truncation and incomplete fixes. 

    & Longer prompts mean more tokens for the model to process, which increases latency and cost. In extreme cases, it might approach context limits and risk truncation. 
    & Simplify prompt and remove redundancy. Use shorter placeholders or variables for lengthy texts if the model can understand them. If many examples are used, see whether fewer achieve similar performance. Monitor token usage and response time to guide prompt-length adjustments.\\
    
    \cmidrule(l){2-4}
    & \textbf{Inefficient few-shot examples.} Providing many or complex examples when a simpler prompt or a fine-tuned model would be more efficient~\cite{logan2021cutting}. \emph{Example}: Using a 10-shot prompt for a task that a zero-shot prompt could handle with minor instruction adjustments. 
    & Unnecessary examples bloat the prompt, again incurring speed and cost penalties. They may confuse the model and also increase performance risk if any example isn’t perfectly aligned with the task. 
    & Use the minimum effective number of examples (i.e., shots). Prefer high-level instructions or simpler demonstrations over exhaustive ones. Evaluate if a specialized model can do the task without heavy prompting. For frequent tasks, consider fine-tuning a model instead of many-shot prompts each time.\\
    
    \cmidrule(l){2-4}
    & \textbf{No prompt caching/reuse.} Re-generating identical prompt segments for each request wastes computation. Identical prefixes get reprocessed every time~\cite{zhu2024efficient}. \emph{Example}: AWS reports up to 85\% lower latency and 90\% lower cost by caching frequent prompts~\cite{aws2024}. 
    & Repeatedly processing the same instructions inflates compute time. Cache hits occur when the input prefix matches exactly; static content at the prompt’s start is ideal for caching. 
    & Implement prompt caching: break the prompt into static and dynamic parts. Place stable sections (guidelines, system instructions) at the beginning so APIs can reuse cached embeddings or KV states. Use memorization of prompt embeddings for repeated use. \\
    
    \cmidrule(l){2-4}
    
    & \textbf{Unbounded output.} The prompt does not constrain answer length or detail. Without explicit limits (e.g., “answer in one paragraph”), the model may generate excessively long responses~\cite{nayab2024concise}. \emph{Example:} When asked to “summarize the codebase,” the model outputs a 10,000-token explanation with redundant details, exceeding API limits and causing downstream truncation. 

    & Long outputs increase generation time and costs linearly. In interactive systems, this hurts responsiveness and may exceed UI or downstream limits. 
    & Constrain output length and scope in the prompt. Use the API’s max\_tokens setting to cap outputs. For large tasks, break them into sub-questions or use iterative refinement to keep each answer concise. \\


    \midrule
    
    \multirow{2}{=}{\textbf{Maintainability \& Engineering}}%
    & \textbf{Hard-coded prompts.} Prompt text is embedded directly in code in multiple places, or scattered across the codebase~\cite{xing2023prompt}. \emph{Example:} A fix to the “code refactoring” prompt in one file is missed in another file, leading to inconsistent results.
    & Inconsistent behavior and difficult updates. One instance of the prompt might be changed (fixing a bug in one context) while other instances remain outdated, causing divergent outputs. 
    & Centralize prompt management. Use a single source of truth for each prompt (e.g., store prompts in configuration files or constants). Adopt templates where dynamic content is filled in to avoid copy-paste modifications, so updates to a prompt propagate everywhere consistently.\\
    

    \cmidrule(l){2-4}
    
    & \textbf{Insufficient prompt testing.} Prompts are not systematically tested with diverse inputs or evaluation metrics~\cite{xing2023prompt}. \emph{Example:} A prompt works well on “sorting Python lists” but fails when handling nested lists due to lack of test coverage.
    & Undetected defects continue until end-users encounter them. A prompt might work for a few sample queries but fail for edge cases, causing bad outputs in production. Lack of testing also makes it hard to improve prompts confidently. 
    & Develop prompt tests and evals. Use tools to write test cases (input–output expectations) and run them automatically. Leverage frameworks for automated prompt evaluation on benchmark datasets. Add these tests to CI/CD pipelines so any change in the prompt or model triggers them, catching regressions before deployment.\\
    
    
    \cmidrule(l){2-4}
    
    & \textbf{Poor documentation.} The prompt’s purpose or intricacies are not documented for future maintainers~\cite{chen2025promptware}. \emph{Example:} No comments explain why certain instructions exist.
    & New developers or team members may not understand why the prompt is written a certain way. They might remove what seem like odd phrases or formatting, unknowingly re-introducing defects the prompt was crafted to avoid. 
    & Document prompts just as you would document code. Add comments inside the prompt string (if supported) or in accompanying docs to explain non-obvious instructions. Record known limitations or work-arounds so institutional knowledge about prompt quirks is preserved.\\
    
    \cmidrule(l){2-4}
    & \textbf{Security/safety review gaps.} Prompt design is not examined for potential abuse cases (injection, leaking secrets, etc.)~\cite{yi2025benchmarking}. \emph{Example:} Prompt accidentally exposes API keys.
    & The system might pass initial tests but be vulnerable in real-world use. For instance, a prompt may inadvertently expose an API key or fail to handle malicious input, leading to a security incident. 
    & Incorporate prompt review into the development lifecycle. Use a checklist for issues such as injection paths, sensitive-data handling, and policy compliance. Leverage tools with \emph{red-teaming} modes to scan for vulnerabilities\cite{derczynski2024garak}. Perform security testing on prompts much like pen-testing an application.\\
    
    \cmidrule(l){2-4}
    & \textbf{Integration mismatch.} This subtype assumes the prompt already specifies an explicit output contract. A mismatch occurs when the model’s response violates that contract, or when the stated contract is not aligned with the downstream parser or UI~\cite{shao2024llms}. \emph{Example:} The prompt requires JSON with fields \texttt{category} and \texttt{defect}, but the model returns \texttt{defects} or extra fields; or the prompt adopts \texttt{snake\_case} while the parser expects \texttt{camelCase}.

    & Even if semantic content is correct, incompatible formatting can cause crashes or silent errors (e.g., JSON decode errors, missing fields). For instance, an LLM might omit expected delimiters or reorder list items, breaking subsequent processing. 
    & Enforce structured output: instruct the model to follow a schema (e.g., JSON, CSV) and validate it in code. Whenever prompts change, update parsing logic accordingly. Test end-to-end integration so any mismatch (format, vocabulary, ordering) is caught before release. \\

\end{longtable}
\end{center}

\section{Discussion}
Prompt defects lie between the written instruction, the prompt, and the system that runs it, the model and its runtime. To reason about these defects, we should separate the two sources of failure. One source is the prompt itself (e.g., ambiguity, missing constraints, or poor structure). The other source is the model or runtime (e.g., limited context length, weak instruction following, or a tendency to hallucinate). We propose an operational view: A defect is a failure mode observed in a specified deployment context (that is, a particular model family, context budget, decoding settings, and acceptance criteria). This makes it natural to use two types of check. Prompt-level checks examine the instruction as an artifact (is it clear, internally consistent, and does it require a specific output format). Model-level checks run the prompt on representative target runtimes and input distributions to see how often and in what ways it fails (e.g., dropping earlier constraints or producing hallucinated facts). Because model capabilities change over time (longer context windows, better instruction following, retrieval integration), the relative importance of different defect types will shift; the taxonomy should therefore be versioned and tied to the contexts in which it was measured. 

\section{Conclusion and Future Work}
\label{sec:conclusion}

In this paper, we presented the first systematic taxonomy of prompt defects in LLM systems, organizing recurring failure modes into six major dimensions and multiple fine-grained subtypes. This taxonomy provides a unified conceptual framework for understanding how prompts fail, why these failures occur, and how they affect downstream LLM-driven applications. By linking each defect type to its practical impact and potential mitigation strategies, our work contributes to the establishment of engineering-oriented methodologies for prompt development.

The taxonomy also highlights open challenges in prompt engineering and LLM-based software systems. One promising direction is the development of automated tools for detecting and repairing prompt defects. Such tools could combine static or dynamic prompt analysis with LLM-based self-repair mechanisms to reduce manual effort and improve system reliability. Another important direction involves building standardized benchmarks for evaluating prompt robustness and correctness under diverse conditions. These benchmarks would enable reproducible comparisons across different defect detection and mitigation techniques. Finally, future work should explore human-centered prompt engineering by integrating usability studies and human-in-the-loop feedback into prompt design workflows. Understanding how users formulate prompts and interact with models will be critical for improving both prompt effectiveness and overall system reliability.

By addressing these challenges, we aim to move from ad-hoc prompt crafting toward principled, engineering-driven methodologies. Ultimately, our goal is to enable LLM-powered systems that are more robust, trustworthy, and maintainable, ensuring that prompt engineering matures into a disciplined and reliable practice.

\bibliographystyle{ACM-Reference-Format}
\bibliography{sample-base}

\end{document}